\documentclass[aps,prd,amssymb,cite,
amsfonts,epsf,preprintnumbers,nofootinbib,superscriptaddress]{revtex4}
\usepackage{graphics,epsfig}
\usepackage{graphicx}
\usepackage{dcolumn}
\usepackage{epstopdf}
\newcommand{\be}{\begin{equation}}
\newcommand{\ee}{\end{equation}}
\newcommand{\beq}{\begin{equation}}
\newcommand{\eeq}{\end{equation}}
\newcommand{\bea}{\begin{eqnarray}}
\newcommand{\eea}{\end{eqnarray}}
\newcommand{\nn}{\nonumber}
\def\be{\begin{equation}}
\def\ee{\end{equation}}
\def\ba{\begin{eqnarray}}
\def\ea{\end{eqnarray}}

\begin{document}
\title{  Anti-Evaporation of  Bardeen de-Sitter Black Holes }

\author{Dharm Veer Singh} 
\email{veerdsingh@gmail.com}

 \author{Naveen K. Singh} 
\email{naveen.nkumars@gmail.com}

\affiliation{Centre for Theoretical Physics,\\
 Jamia Millia Islamia,  New Delhi 110025
 India. \vspace{15mm}}

\begin{abstract}
\vspace{5mm}
\noindent  In this paper, we discuss the possibility of the anti-evaporation of degenerate Bardeen de-Sitter black hole. We solve 
the perturbation equations around the Nariai space-time. The solution of one of the perturbations related to the horizon size demonstrates that
horizon of such black hole is constant. The other perturbation is also found to be stable.  We further study thermodynamical  properties of
such black holes. We observe  double phase transition at the Nariai limit. 
\end{abstract}
\vspace{10mm}

\maketitle

\section{\label{sec:level1}Introduction}
Anti-evaporation  is an essential phenomenon which could survive the primordial black holes till the current era of Universe. One of
the importance of primordial black holes is that it may convey the information of early Universe. It may also be a candidate for 
the dark matter of Universe \cite{Ivanov:1994pa,Blais:2002nd,Chavda:2002cj}. There are some theories, which support the formation
of the black hole during the early Universe, e.g., initial density perturbation \cite{hawking1,hawkingCarr} and non-linear metric 
perturbation  \cite{Bullock:1996at,Ivanov:1997ia,Bullock:1998mi}, the evolution of gravitational bound objects during the early 
time \cite{Kalashnikov}.\\

The well-known phenomenon of evaporation of black hole was introduced by Hawking \cite{Hawking:1974sw}. In this process the black hole 
horizon shrinks. However, on contrary to this, Hawking and Bousso have  described the possibility of 
anti-evaporation \cite{Bousso97} of a black hole due to quantum  process, which appears in the 
limit where the cosmological and event horizon coincide, i.e., in Nariai space-time \cite{Nariai1, Nariai2}. 
Anti-evaporation is relevant in multiple horizon Nariai black hole and also explained by common grand unification theories 
\cite{Elizalde:1999dw,Bytsenko:1998md}. Primordial black hole can  be created during inflation by pair creation or by quantum
fluctuations, however they decay quickly due to Hawking radiation.
In the anti-evaporation process, the
radius of the black hole horizon increases \cite {Nojiri:1998ph, Buric:2000cj}. If there is only hawking radiation, then the primordial black hole
can not be observed. However, anti-evaporation could prevent the energy of black hole so that primordial black hole may survive even
in the current era of the Universe.\\

Usually, we require to include the quantum correction from the matters to explain the anti-evaporation. However, the anti-evaporation 
may be possible  even on the classical level.  The former is studied 
in Refs. \cite{Bousso97, Niemeyer:2000nq,Nojiri:1998ue}, where the calculation is done by using two-dimensional one-loop effective action 
with the s-wave approximation. Furthermore, the four dimensional conformal anomaly also indicates the possibility 
\cite{Niemeyer:2000nq,Nojiri:1998ph, Nojiri:1998ue,Nojiri:2000ja}. Latter one,
i.e., the anti-evaporation due to the classical effect, can be possible in $F(R)$  gravity \cite{Nojiri13, Addaz}, mimetic $F(R)$ gravity 
\cite{Oikonomou:2015lgy}. Therefore, Anti-evaporation can be arisen due to both quantum  as well as classical
effects. Some progress in the direction of anti-evaporation have also been made in theory of bigravity \cite{taishi}.  We may think
of anti-evaporation in Bardeen de-Sitter black hole, since it has cosmological horizon and it has non-linear source. In this paper, we will 
analyze the perturbation around the Nariai space-time and discuss the possibility of anti-evaporation at the classical level in
this black hole.  \\

Bardeen black hole is a regular black hole, i.e., it does not have singularity inside the horizon. It was first introduced
 by Bardeen \cite{bardeen1968}. Later on, in the Ref. \cite{gar1}, authors explained it by assuming non-linear Electrodynamics. Here, they introduced self-gravitating
 magnetic field by a monopole charge and developed the corresponding theory. Some more works on Bardeen black hole have been done in Refs. 
 \cite{sar,fernando1,lemos,ulhoa}.  Since the current cosmological data proves that the universe is accelerating,  some of the properties of Bardeen black hole were 
 studied in Ref. \cite{b1.1} by   adding the cosmological constant. The cosmological constant  is one of the candidates for the dark energy 
 which gives rise the acceleration  in the current Universe.  The cosmological constant introduces a cosmological horizon. Therefore, it is
 interesting to test if anti-evaporation is possible in this black hole.
 We will discuss mainly anti-evaporation of Bardeen de-Sitter black hole and for completeness, we also study the 
 instability  of perturbation and thermodynamical properties of this black hole in the limiting case of Nariai space-time. \\
 
%

The paper is organized as follow: In Sec. \ref{sec:level2}, we briefly review a mechanism for anti-evaporation in $f(R)-$gravity. We 
discuss the solution of Bardeen de-sitter  black hole in Sec. \ref{sec:level3}. In Sec. \ref{sec:level4}, we study the
anti-evaporation and instability of the perturbations for the  Bardeen-Nariai black hole in the presence of cosmological constant.
Its thermodynamics  is presented in Sec. \ref{sec:level5.3}. The conclusion and future direction are given in Sec. \ref{sec:level6}.
\section {Anti-Evaporation in $F(R)$ gravity}
\label{sec:level2}
In this section, we review the  formalism of anti-evaporation in a black hole at classical level by  Odintsov and Nojiri  in $f(R)$ gravity
\cite{Nojiri13} which was generalized from a theory of anti-evaporation by Bousso and Hawking \cite{Bousso97}.  In $f(R)$ gravity
the anti-evaporation is possible due to the purely classical effect.  The anti-evaporation occurs when cosmological and event
horizon coincide and the corresponding solution is called the Nariai solution. Let us consider the action of $f(R)$ theory is
\be
S=\frac{1}{16\pi G}\int d^4x\sqrt{-g} f(R) +S_m,
\label{act1}
\ee
where $G$ is Newton constant and R is Ricci scalar. The field equation corresponding to the action (\ref{act1}) is
\be
f^{\prime}(R)R_{\mu\nu}-\frac{1}{2}f(R)g_{\mu\nu}-\nabla_{\mu}\nabla_{\nu}f^{\prime}(R)+g_{\mu\nu}\Box f^{\prime}(R)=8\pi G T_{\mu\nu},
\ee
where $T_{\mu\nu}$ is the energy momentum tensor. Assuming no matter ($T_{\mu\nu}=0$) and Ricci tensor is covariantly constant, i.e.,
proportional to the metric $g_{\mu\nu}$, the field equation reduce to 
\be
 f(R)-\frac{1}{2} Rf^{\prime}(R)=0.
\label{act3}
\ee
The Eq. (\ref{act3}) admits Nariai space-time  solution, where the line element is given by
\be
ds^2=\frac{1}{\Lambda^2}\frac{1}{\cosh^2 x}(d\tau^2-dx^2)+ r_0^2 d\Omega^2,
\label{nar1}
\ee
where we need to  introduce the new coordinates $\tau$ and $x $ related to $t$ and $r$ via 
$t=2 r_0^2 \tau/[(1-R_0r_0^2/2) \epsilon]$ and $r=r_0+\frac{\epsilon}{2}(1+\tanh x)$. There are two horizons at $r_0$ and $r_1$ separated by a distance
$\epsilon\rightarrow 0$ ($r_1=r_0+\epsilon$) for Nariai space time. $R_0$ is constant Ricci scalar.  For $R_0=0$, $\Lambda$ becomes $1/r_{0}$.
In order to understand the dynamics of  the Nariai  space-time, we need to perturb the metric  (\ref{nar1}). The perturbed metric is
\be
ds^2=\frac{e^{2\rho(x,\tau)}}{\Lambda^2}(d\tau^2-dx^2)+ \frac{ e^{-2\phi(x,\tau)}}{\Lambda^2}d\Omega^2,
\label{nar11}
\ee
with
\ba
\rho=-\ln(\Lambda\,\cosh x)+\delta\rho,\qquad\qquad \phi=\ln\, \Lambda+\delta\,\phi.
\ea
Substituting these expressions into field equations, a set of equations in perturbations $\delta\rho$ and $\delta\phi$ is obtained. 
By considering  $F''(R_0)\neq 0$, a unique solution is obtained which is given by,
\be
\delta \rho=\rho_0\,\cosh \omega \tau\,\cosh^\beta x,\qquad\qquad \delta \phi= \phi_0\,\cosh \omega \tau\,\cosh^\beta x,
\ee
where $\omega$ and $\beta$ are real parameter related by the field equation and horizon radius $r_h$. The horizon of the black hole can be defined by the relation
\be
g^{\mu\nu}\nabla_{\mu}\phi\nabla_{\nu}\phi=0.
\ee
By assuming $\delta \phi= \phi_0\,\cosh \omega t\,\cosh^\beta x$, we have
\be
\omega^2\tanh^2\omega \tau=\beta^2\tanh^2x,
\ee
because of $\omega^2=\beta^2$ \cite{Nojiri13},  $\delta\phi=\delta\phi_h=\phi_0\cosh^2\beta t$. Eq. (\ref{nar11}) shows that $e^{-\phi}/\Lambda$ can be regarded as a radius coordinate, one may define the radius of the horizon using the Eq. (\ref{nar1}) by  
\be
r_h=\frac{e^{-\delta\phi_h}}{\Lambda}= \frac{e^{-\phi_0\, \cosh^2 \omega \tau}}{\Lambda}.
\ee
If $\phi_0< 0$, the horizon radius of the black hole increases and  the anti-evaporation occurs corresponding to the increasing 
absolute value of $|\delta \phi|$.  Using this formalism, we study the anti-evaporation problem in Bardeen de-Sitter  Black Hole. If we
include the quantum correction the evaporation may be possible. However, we will consider only the classical phenomenon in this paper.

\section{Bardeen de-Sitter  Black Hole}
\label{sec:level3}
 The  action of Einstein gravity in the presence of nonlinear electromagnetic field and cosmological constant is 
\be
S=\frac{1}{16\pi}\int d^4x \sqrt{-g}\left[R-2\Lambda-4{\cal L}(F)\right]. \label{action}
\end{equation}
Here $R$ is Ricci scalar,  $\Lambda$ is a  cosmological constant and ${\cal L}(F)$ is the Lagrangian of nonlinear
electromagnetic field   and has the following form,
\be
{\cal L}(F)=\frac{3}{2 s\, g^2}\left(\frac{\sqrt{2g^2F}}{1+\sqrt{2g^2F}}\right)^{\frac{5}{2}},
\ee
where $F=F_{\mu\nu}F^{\mu\nu}$, $F_{\mu\nu}$ is the electromagnetic field tensor, g and M are the magnetic monopole charge and mass of
black hole, related by equation $s = |g|/2M$. The equations of motion can be obtained by varying the action (\ref{action}) with respect to the metric $g_{\mu\nu}$, the
maxwell field $A_{\mu}$ respectively. These equations are given as,
\ba
&&G_{\mu\nu}+\Lambda g_{\mu\nu}=T_{\mu\nu},\nn
\ea
\ba
\nabla_{\mu}\left(\frac{\partial{ \cal L}(F)}{\partial F}F^{\mu\nu}\right)=0 \qquad\text{and}\qquad \nabla_{\mu}(*F^{\mu\nu})=0.
\label{eom}
\ea 
and the energy momentum tensor turns out to be,
\be
T_{\mu\nu}=2\left(\frac{\partial{ \cal L}(F)}{\partial F}F_{\mu\lambda}F^{\lambda}_{\nu}-g_{\mu\nu}{\cal L}(F)\right).
\label{em}
\ee
In order to find the static spherical symmetric black hole solution, we consider the following line element
\be
ds^2=f{(r)}dt^2- \frac{1}{f(r)}dr^2 - r^2 d\Omega^2
\ee
with
\be
f(r)=\left(1-\frac{2m(r)}{r}\right).
\ee
Following the ansatz for Maxwell equation
$F_{\mu\nu}= 2 \delta^{\theta}_{[\mu}\delta^{\phi}_{\nu]} Z(r,\theta)$ \cite{gar1} which leads only non zero components
$F_{\theta\phi}=-F_{\phi\theta}=g\sin \theta$ and on solving the equations (\ref{eom}) with energy momentum tensor (\ref{em})  we
obtain the  following solution,
\be
f(r)=1-\frac{2mr^2}{(r^2+g^2)^{3/2}}-\frac{\Lambda r^2}{3}.
\label{bar1}
\ee 
This is the solution of regular de-Sitter black hole called the Bardeen de-Sitter black hole \cite{b1.1}. The Nariai Solution is obtained 
when the cosmological and event horizon coincide. The detail study of horizons is discussed in Ref. \cite{b1.1}. The horizon of the
Nariai space-time solution (one of the solution for extreme black hole) is given by,

\be
r_{ex}^2=\frac{1+\sqrt{1-8\Lambda g^2}}{2\Lambda}
\ee
with associated  mass of the black hole,
\be
M_{ex}=\frac{\left(g^2+r_{ex}^2\right)^{5/2}\Lambda}{3(r^2_{ex}-2g^2)}.
\ee
The Nariai solution  exists only if $8\Lambda g^2<1$. Near the Nariai space-time, we may write the approximated form of the function $f(r)$ as 
\be
f(r)=\frac{f''(r_{ex})}{2}(r-r_c)(r-r_h),
\ee 
where, $r_c$ and $r_h$ are cosmological horizon and event horizon respectively. Now, we  transform the old coordinates $r$ and $t$ to new coordinates $\chi$ and $\psi$ as follows
\be
r=r_{ex}+\epsilon \,\cos\chi \qquad\text{and}\qquad t=\frac{2\psi}{\epsilon f''(r_{ex})},
\ee
where $\epsilon$ is a parameter which differentiate two horizons. For $\chi=0$ and $\chi=\pi$, we get $r=r_{ex}+\epsilon=r_c$ and $r=r_{ex}-\epsilon=r_h$ respectively.
 We note that for $\epsilon\rightarrow 0$ both horizons coincide. In this limiting case, 
 the line element  of  Bardeen de-Sitter black holes in new coordinates becomes
\be
ds^2=\frac{2}{f''(r_{ex})}\left(d\chi^2-\sin^2\chi\,d\psi^2\right)-r^2_{ex}d\Omega^2.
\label{nar2}
\ee
In this space-time $f''(r_{ex})$ is given by
\begin{equation}
f''(r_{ex} )= \frac{2 r_{ex}^2 \Lambda (r_{ex}^2-4 g^2)}{2 g^4 + g^2 r_{ex}^2 - r_{ex}^4}.
\end{equation}
The line element (\ref{nar2}) is similar as in the case of the Nariai space-time
\be
ds^2=\frac{1}{\Lambda^2\, \cosh^2 x}(d\psi^2-d \chi^2)-\frac{r^2}{\Lambda^2} d\Omega^2,
\ee
with a difference of  the signatures for $\psi$ and $\chi$. Next we study the perturbation around the solution for the Nariai space-time.
\section{\label{sec:level4}Anti-Evaporation}
In this section, we study the   anti-evaporation of Bardeen de-Sitter black hole in the limit of Nariai space-time. The solution is obtained
from the equation of motion  at the linear order of the perturbation theory. 
 In addition, we will also discuss the instabilities 
 of perturbation of Bardeen de-Sitter black hole in the Nariai limit.
\subsection{\label{sec:level4.1}Perturbation}
 In general, to know how the horizon evolves, we perturb the space-time as follows,
 \begin{eqnarray}
  ds^2 =  \frac{2}{f''(r_{ex})} \left( - \sin^2(\chi) d\psi^2 + d\chi^2\right) e^{2 \delta \rho(\psi,\chi)}
  - r_{ex}^2    e^{-2 \delta \phi(\psi,\chi)} d\Omega^2.
 \end{eqnarray}
However, the off-diagonal Einstein's equation demands that the spatial part of perturbations should be proportional to $\sin(\chi)$.
We may consider $\delta \rho(\psi,\chi) = A f_{1} (\psi) f_3(\chi)$ and $\delta \phi(\psi,\chi) = B f_{2}( \psi) f_3(\chi)$. Under this assumption,
we find,
 \begin{eqnarray}
  f_3' = f_3\cot{\chi},
 \end{eqnarray}
where prime $'$ denotes the derivative with respect to $\chi$. Integrating,  we get the solution as $f_3= f_{30}\sin{\chi}$ with $f_{30}$ as a constant. Therefore, we choose the forms of perturbations as following
 \begin{eqnarray}
  ds^2 =  \frac{2}{f''(r_{ex})} \left( - \sin^2(\chi) d\psi^2 + d\chi^2\right) e^{2 A f_1( \psi) \sin(\chi)}
  - r_{ex}^2  e^{-2 B f_2( \psi) \sin(\chi)}  d\Omega^2 ,
 \end{eqnarray}
where, $A$ and $B$ are constants. Upto the linear order $\psi-\psi$, $\chi-\chi$ and $\theta-\theta$ part of  Einstein equation are given by
\begin{eqnarray}
 \frac{2 \sin(\chi)}{f''(r_{ex}) r_{ex}^2(g^2+r_{ex}^2)^3}\Bigg[f_1(\psi)\Big[2 A (g^2 + r_{ex}^2)(1- r_{ex}^2 \Lambda)\left(r_{ex}^4+ g^4 
 + 2 g^2 r_{ex}^2 \left(1- \frac{3 M}{\sqrt{(g^2+r_{ex}^2)}(1- r_{ex}^2 \Lambda)}\right)\right)\Big] \nonumber \\
 + f_2( \psi) B [2 + f''(r_{ex}) r_{ex}^2]\left(g^6 + 3 g^4 r_{ex}^2 
 + r_{ex}^6 + 3 g^2 r_{ex}^4 \left(1 - \frac{10 M}{\sqrt{(g^2+r_{ex}^2)}[2 + f''(r_{ex}) r_{ex}^2]}\right)\right)\Bigg]=0, \label{einstein1}
\end{eqnarray}

\begin{eqnarray}
 -\frac{2 \sin(\chi)}{f''(r_{ex}) r_{ex}^2(g^2+r_{ex}^2)^3}\Bigg[f_1(\psi)\Big[2 A (g^2 + r_{ex}^2)(1 - r_{ex}^2 \Lambda)\left(r_{ex}^4+ g^4 
 + 2 g^2 r_{ex}^2 \left(1- \frac{3 M}{\sqrt{(g^2+r_{ex}^2)}(1- r_{ex}^2 \Lambda)}\right)\right)\Big] \nonumber \\
 + f_2(\psi) B [2 + f''(r_{ex}) r_{ex}^2]\left(g^6 + 3 g^4 r_{ex}^2 
 + r_{ex}^6 + 3 g^2 r_{ex}^4 \left(1 - \frac{10 M}{\sqrt{(g^2+r_{ex}^2)}[2 + f''(r_{ex}) r_{ex}^2]}\right)\right) \nonumber \\
 - f''(r_{ex}) B r_{ex}^2  \csc^2(\chi)(g^2 + r_{ex}^2)^3 \Big[f_2(\psi) -  \ddot{f_2}{(\psi})\Big]\Bigg]=0, \label{einstein2}
\end{eqnarray}

\begin{eqnarray}
 \frac{\sin(\chi)}{2}\Bigg[A f''(r_{ex}) \csc^2(\chi) f_1(\psi) + B f_2( \psi)\Big[4 f''(r_{ex})
  + 2 \left( \frac{3 g^2 M}{(g^2 + r_{ex}^2)^{9/2}} (4 g^4 - 22 g^2 r_{ex}^2 + 9 r_{ex}^4) + 2 \Lambda\right) \nonumber \\ - f''(r_{ex}) \csc^2(\chi)\Big] 
  + f''(r_{ex}) \csc^2(\chi) \Big[- A  \ddot{f_1}(\psi)+ B  \ddot{f_2} (\psi)\Big]\Bigg]=0 .  \label{einstein3}
\end{eqnarray}
Adding Eq. (\ref{einstein1}) and Eq. (\ref{einstein2}), we have
\begin{eqnarray}
f_2(\psi) -  \ddot{f_2}{(\psi})=0. \label{dif_eqn}
\end{eqnarray}
The solution of Eq. (\ref{dif_eqn}) is given by,
 \begin{eqnarray}
  f_2= C_1 e^{\psi} + C_2 e^{-\psi}.
 \end{eqnarray}
 where, $C_1$ and $C_2$ are constants. Therefore, the solution for $\delta \phi$ is given by
\begin{eqnarray}
\delta \phi =  B \sin(\chi)\Big[C_1 e^{\psi} + C_2 e^{-\psi}\Big]. \label{analy_phi}
 \end{eqnarray}
One can write the above solution in  more simpler form by using the condition at the horizon,
\begin{eqnarray}
 g^{\mu\nu} \nabla_{\mu} \delta \phi \nabla_{\nu} \delta \phi =0, \label{hori_cond}
\end{eqnarray}
which provides a relation between the coordinate $\chi$ and $\psi$ as follows,
\begin{eqnarray}
 \cos{\chi} = \frac{C_1 e^{\psi}- C_2 e^{-\psi}}{C_1 e^{\psi}+ C_2 e^{-\psi}}
\end{eqnarray}
Or,
\begin{eqnarray}
 \sin{\chi}= \frac{2 \sqrt{C_1 C_2}}{C_1 e^{\psi}+ C_2 e^{-\psi}}. \label{sinchi}
\end{eqnarray}
Thus, the solution now becomes,
\begin{eqnarray}
 \delta \phi = 2 B \sqrt{C_1 C_2}. \label{analy_phi_new}
\end{eqnarray}
 We notice that the perturbation $\delta \phi$ becomes constant with respect to $\psi$. The perpurbation might have increasing behavior with
 respect to $\psi$ (see Eq. \ref{analy_phi} ). However, at horizon,  we have a constant soution for the perturbation which is due to
 Eq. (\ref{hori_cond}).  In the following subsection (\ref{sec:level3.2}), we discuss about the size of horizon and other perturbation $\delta \rho$.

\subsection{Horizon Tracing and instability}\label{sec:level3.2}

As stated above, the perturbation $\delta \phi$ becomes  constant at the horizon. Now the  horizon $r_h$ is given by
\begin{eqnarray}
 r_h = r_{ex}e^{-4 B \sqrt{C_1 C_2}}.
\end{eqnarray}
Here, we note that  we have constant solution for horizon which proves that the black hole at classical level does not (anti-)evaporate.
We now look for the other  perturbation part $\delta\rho$ by using Eq. (\ref{einstein3}). The Eq. (\ref{einstein3}), on arranging, may be written as
\begin{eqnarray}
  \frac{2 B}{A f''(r_{ex})} \sin^2{\chi} \Big[2 f''(r_{ex})+ \left( \frac{3 g^2 M}{(g^2 + r_{ex}^2)^{9/2}} (4 g^4 - 22 g^2 r_{ex}^2 + 9 r_{ex}^4) + 2 \Lambda\right)\Big] f_2
 + \frac{B}{A} \left(\ddot{f_2}-f_2\right) -  \ddot{f_1} + f_1=0,
\end{eqnarray}
 Simplifying, we obtain,

\begin{eqnarray}
  \frac{ B}{A} \sin^2{\chi} \Big[\frac{12 g^4 - 21 g^2 r_{ex}^2 + 2 r_{ex}^4}{(r_{ex}^2 - 4 g^2)(g^2+ r_{ex}^2)} \Big] f_2
 + \frac{B}{A} \left(\ddot{f_2}-f_2\right) -  \ddot{f_1} + f_1=0, \label{einstein3_si}
\end{eqnarray}
Using Eqs. (\ref{dif_eqn}) and (\ref{sinchi}), the above Eq. (\ref{einstein3_si}) turns out to be,
\begin{eqnarray}
  \frac{ 2 B \sqrt{C_1 C_2}}{A}\sin{\chi} \Big[\frac{12 g^4 - 21 g^2 r_{ex}^2 + 2 r_{ex}^4}{(r_{ex}^2 - 4 g^2)(g^2+ r_{ex}^2)} \Big] 
  - \ddot{f_1} + f_1=0, \label{einstein3_final}
\end{eqnarray}
Or,
\begin{eqnarray}
 \ddot{f_1} - f_1 = \gamma ; \ \ \ \gamma =  \frac{ 2 B \sqrt{C_1 C_2}}{A}\sin{\chi} \Big[\frac{12 g^4 - 21 g^2 r_{ex}^2 + 2 r_{ex}^4}{(r_{ex}^2 - 4 g^2)(g^2+ r_{ex}^2)} \Big]
\end{eqnarray}
Thus, in the limit $\psi \rightarrow \infty$, where $\gamma \rightarrow 0$, the solution of  Eq. (\ref{einstein3_final}) can be written as
\begin{eqnarray}
 f_1 = C_3 e^{\psi} + C_4 e^{-\psi} ,
\end{eqnarray}
where, $C_3$ and $C_4$ are constants. The above solution provides $\delta \rho$ as
\begin{eqnarray}
 \delta \rho  &=& \frac{2 A \sqrt{C_1 C_2} }{(C_1 e^{\psi} + C_2 e^{-\psi})} \Big[C_3 e^{\psi} + C_4 e^{-\psi} \Big] \nonumber \\
 &\sim& 2 A C_3 \sqrt{\frac{C_2}{C_1}} \ \  (\psi \rightarrow \infty). \label{analy_rho}
\end{eqnarray}


We notice that the Nariai space-time arises in the case where condition $r_{ex}^2= \left(1 + \sqrt{1-8 g^2 \Lambda}\right)/(2 \Lambda)$ is 
satisfied, i.e., where event and cosmological horizon coincide. The perturbation equations also depend on the all the constraints equations which are due to 
this special case. Nariai space-time enforces to the perturbations $\delta \phi$ and $\delta \rho$ to be unstable in general. However,
at the horizon, due the condition (\ref{hori_cond}) the perturbations $\delta \phi$, $\delta \rho$ are stable.
The growing and the decreasing solutions for horizon represent anti-evaporation and evaporation respectively.
We obtain constant horizon solution for Bardeen black hole in this space-time which explain Bardeen de-Sitter black hole does not 
(anti-)evaporate classically. We can compare the solution with solution obtained in $F(R)$-theory \cite{Nojiri13}. In $F(R)$-theory,
we can get  evolution of horizon despite the presence of involved constraint relations. However, we obtained different solution from 
f(R)-theory. The constant solution is obtained, since non-linear Electrodynamic term does not involve in the dynamics. The perturbation $\delta \rho$ has also
similar behavior. It has a finite value for large $\psi$.  We may  think of standard Hawking radiation effect as well as
anti-evaporation effect due to the quantum effect. If anti-evaporation would be effective then it might prolong the lifetime of primordial black hole. The analytical
solution of perturbation $\delta \phi$ and $\delta \rho$ are given in Eq. (\ref{analy_phi_new}) in Eq. (\ref{analy_rho}) respectively.

\section{\label{sec:level5.3}Thermodynamics}
\begin{figure}[h]
\begin{tabular}{c c c c} 
\includegraphics[width=0.47\linewidth]{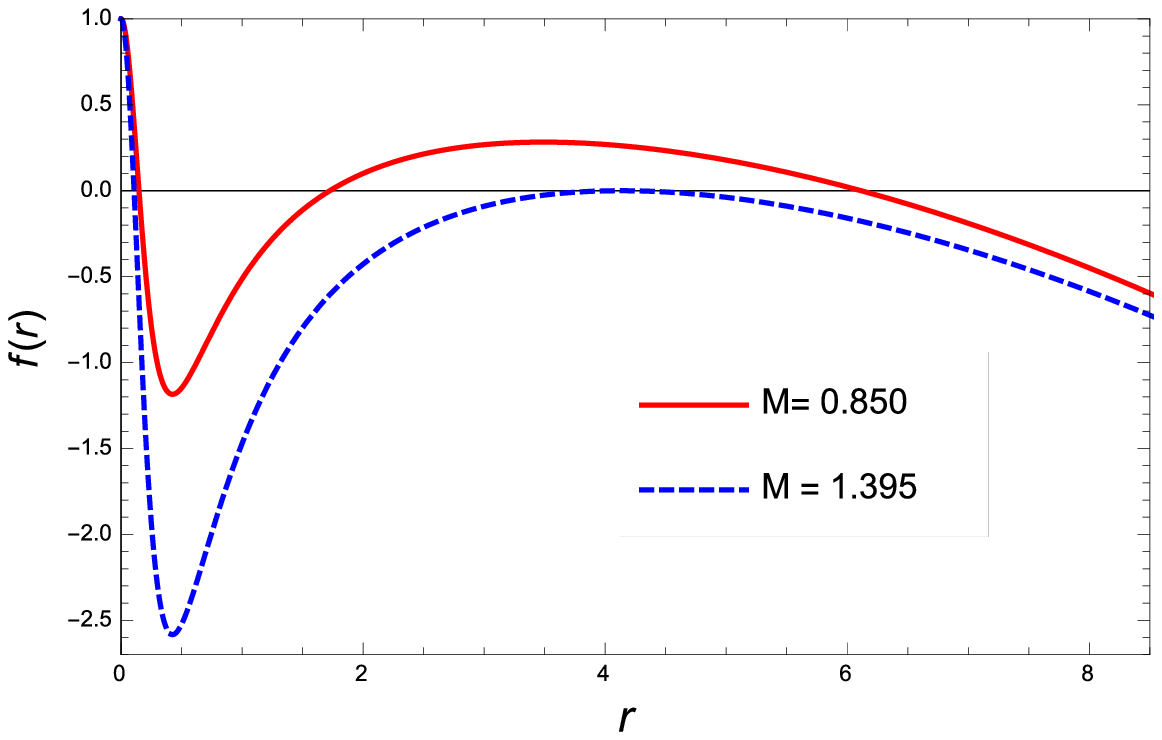}
\includegraphics[width=0.45\linewidth]{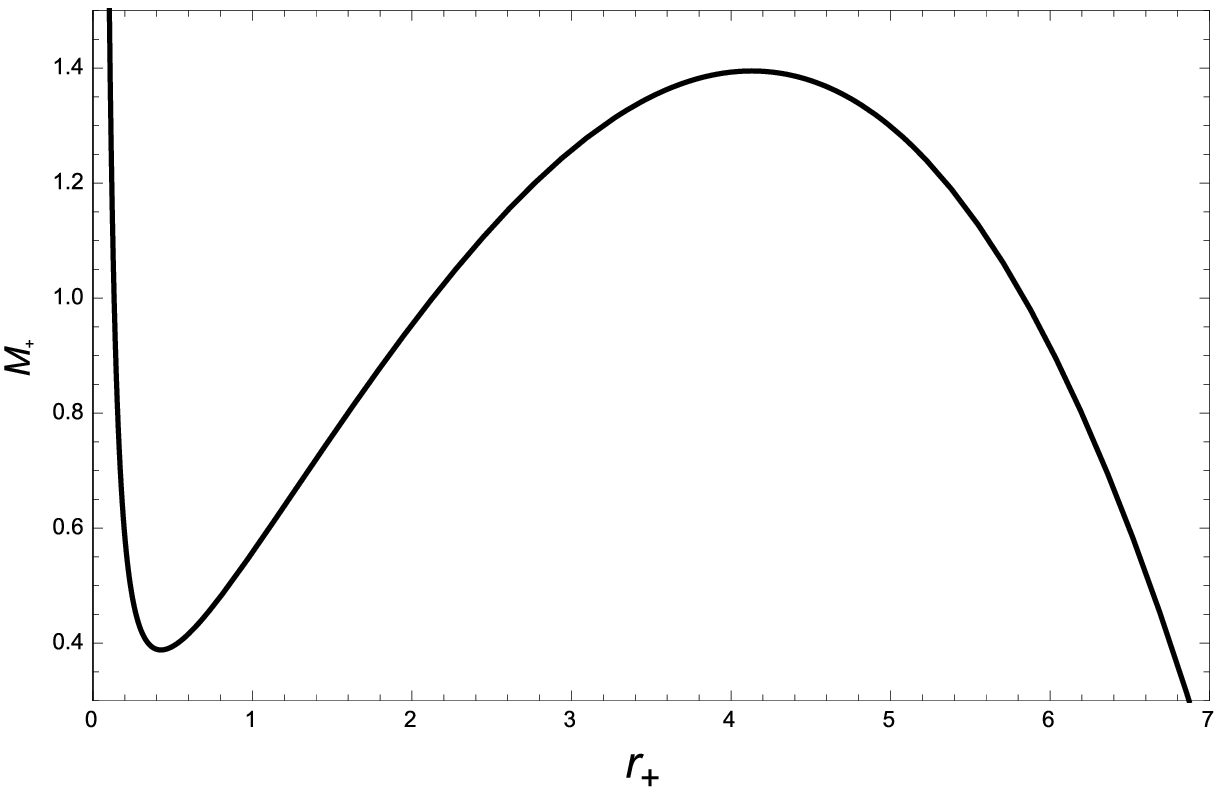}
\end{tabular}
\caption{\label{fig:RK3}The plot of f(r) vs. r is in left panel. The black hole has three horizons viz: inner, outer and cosmological one. The 
mass vs.  horizon radius  with fixed values of $g=0.30$ and $\Lambda=0.058$ is given in the right panel.}
\end{figure}

 In this section, we study the  thermodynamical properties of Bardeen de-Sitter black hole in the limit of Nariai space-time. The black hole 
 is characterized by mass $M$, cosmological constant $\Lambda$ and magnetic monopole charge $g$. The horizon is obtained
 by $f(r)=0$ and plotted in Fig. (\ref{fig:RK3}) with fixed value of the parameters.  The Nariai black hole appears when
 the two horizons have the same size ( cosmological and event horizons). The degenerate horizon appears  at $r_{ex}=4.13$  with a
 set of parameters $M,g$ and $\Lambda$ equal to 1.395, 0.3 and 0.058 respectively. The mass of the black hole is defined as via horizon radius $r_{+}$,
\be
M_{+}=\frac{(g^2 + r_{+}^2)^{3/2} (3 - r_{+}^2 \Lambda)}{6 r_{+}^2}.
\label{mass}
\ee

The size of the black hole varies between the inner horizon ($r_{in}$) and cosmological horizon $r_{c}$. When the size of black hole
approaches to the inner horizon, one would expect that the radiation coming  from the direction of cosmological horizon $r_c$ is negligible and 
the black hole evaporates like Bardeen black hole. However, in  Nariai black hole two horizons will have the same 
size and the temperature, therefore it may anti-evaporate. The temperature ($T=\kappa/2\pi$, where $\kappa$ is the surface gravity $\kappa=f'(r_{+})/2$)  associated
with the black hole solution (\ref{bar1})  reads  
\be
T_{+}=\frac{1}{4\pi}\left[\frac{(r_+^2-2g^2)(3-r_+^2\Lambda)}{3r_+(g^2+r_+^2)}-\frac{2r_+\Lambda}{3}\right].
\label{eqtp}
\ee
 \begin{figure}[h]
\begin{tabular}{c c c c} 
\includegraphics[width=0.45\linewidth]{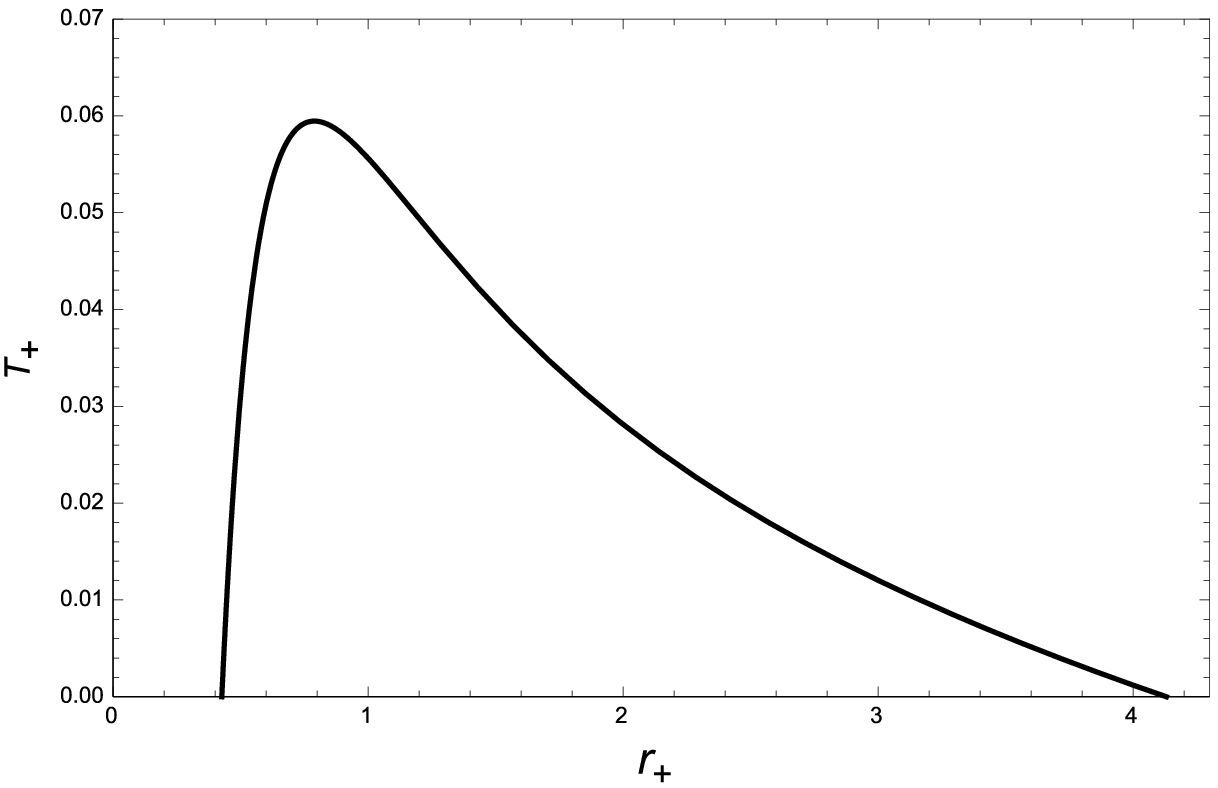}
\includegraphics[width=0.45\linewidth]{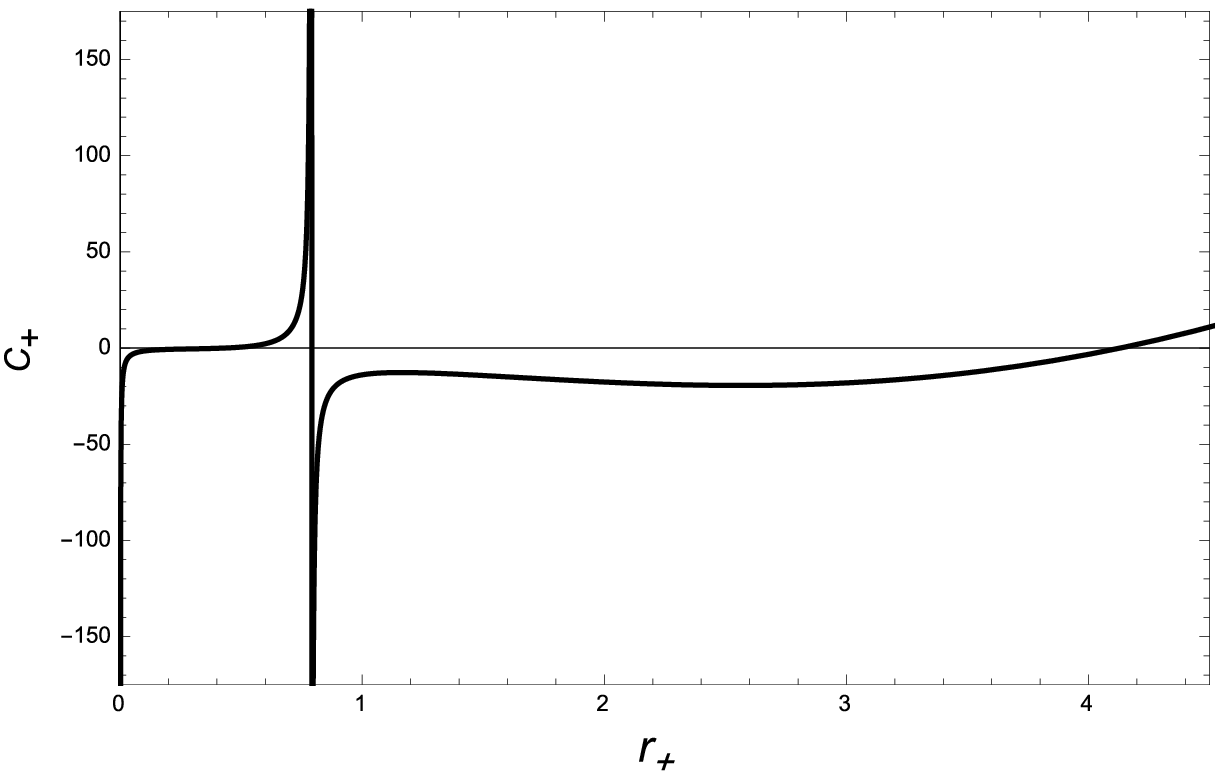}
\end{tabular}
\caption{\label{fig:RK1} Plot of temperature and heat capacity vs.  horizon radius $r_+$ at fixed value of $g=0.30$ and $\Lambda=0.058$ respectively.}
\end{figure}

The temperature reaches at the maximum value at $T_{max}=0.5946$ at $r_+=0.7926$  and vanishes 
at $r_{ex}=4.1303$.  Finally,  we analyze  the thermodynamic stability of  Bardeen de-Sitter black hole. The heat capacity 
is defined as
\be
C_{+}=\frac{\partial M}{\partial T}=\left(\frac{\partial M}{\partial r_+}\right) {\Big/} \left(\frac{\partial r_+}{\partial T}\right).
\label{sh}
\ee
substituting the value of mass (\ref{mass}) and temperature (\ref{eqtp}) into (\ref{sh}), we obtain the specific heat 
\ba
C_{+}=\frac{2 \pi (g^2 + r_+^2)^{5/2} (2 g^2 - r_+^2 + r_+^4 \Lambda)}{-2 g^4 r_+ + r_+^5 + r_+^7 \Lambda + g^2 r_+^3 ( 3 r_+^2 \Lambda-7)}.
\label{eqcp}
\ea

It is difficult to analyze the heat capacity analytically hence we plot it in Fig. (\ref{fig:RK1}) with a particular value of mass $M$,  cosmological 
constant $\Lambda$ and magnetic monopole charge $g$. The thermodynamical stability depends upon the sign of heat 
capacity.  $C_{ex}>0$ and $C_{ex}<0$ correspond the thermodynamical stable and unstable phase respectively. We observed that the 
heat capacity is discontinuous  at inner horizon $r_{in}=0.7926$ at which the  temperature reaches to the maximum value $T_{max}=0.5946$. Hence 
the phase transition occurs from a region with positive heat capacity to  a region of negative heat capacity. We also 
observed that the double phase transition at $r_{ex}=4.1303$, where the temperature vanishes.

\section{conclusion}\label{sec:level6}

The existence and importance of primordial black hole have been studied in the some of the literature as mentioned in the introduction. Anti-evaporation could 
increase the lifetime of such black hole. Since Bardeen de-Sitter black hole is a regular black hole and has cosmological horizon due to the 
cosmological constant term, we studied anti-evaporation for this black hole under the condition where event horizon and cosmological horizon becomes
same. We found that under such condition of Nariai space-time, the size of horizon is constant with time which proves that Bardeen de-Sitter black hole does not
anti-evaporate. There could be possibility of anti-evaporation due to quantum effect or classically if we consider $F(R)$-gravity. One may look for 
anti-evaporation in both scenarios. If anti-evaporation is found under such assumption then  this kind of black hole may live for long time and could be created in the very
early universe. \\

For the comparison to the other recent theories which describe anti-evaporation,
some discussion are in the order. In the Ref. \cite{taishi}, bigravity is tested for the phenomenon of anti-evaporation at classical level and it is
shown that anti-evaporation is not possible in bigravity and it is concluded that it might be possible in $F(R)$-bigravity. Furthermore, in very recent work \cite{Addazi:2017puj}, authors are developing
five-dimensional $F(R)$-theory to explain anti-evaporation of black hole. In $F(R)$-gravity, increasing horizon solution is obtained. In $F(R)$-theory, the anti-evaporation is possible due the complicated nature of gravity with some constraints. However,
in the considered theory, we did not find anti-evaporation due to constant solution for the horizon. This is because of  non-linear  source term 
of electrodynamic is not effective in the dynamical calculation for the perturbation $\delta \phi$. Furthermore,
cosmological constant plays an essential role for the cosmological horizon. Thus it might be interesting if one performs perturbation analysis
 with considered non-linear source term of electrodynamic in $F(R)$-gravity removing the cosmological constant term. \\

We discussed about the perturbations around the Nariai metric.  We have got analytical
solution for both the perturbations. We found that they are stable at horizon. This is an interesting result which motivates for further study
of anti-evaporation in this black hole either considering quantum effect or $F(R)$-gravity at classical level. We, finally, studied the thermodynamical properties of
Bardeen de-Sitter black hole. We observed that the temperature at Nariai-horizon becomes zero and at inner horizon it is maximum. Furthermore,
the heat capacity is discontinuous at the inner horizon and there is double phase transition at Nariai-horizon.

\begin{acknowledgements}
Dharm Veer Singh  and Naveen K. Singh acknowledge the University Grant Commission, India, for financial support through the  D. S. Kothari Post Doctoral Fellowship (Grant No.: BSR/2015-16/PH/0014) and (Grant No.: BSR/2014-15/PH/0034). 
\end{acknowledgements}

\end{document}